\documentstyle[aaspp4,12pt]{article}
\def\Mesz{M\'esz\'aros~}

\def\beq{\begin{equation}}
\def\enq{\end{equation}}
\def\bea{\begin{eqnarray}}
\def\ena{\end{eqnarray}}
\def\bec{\begin{center}}
\def\enc{\end{center}}
\def\etal{{\it et al.}}

\def\tl{\tableline}

\begin{document}

\title{ Impulsive and Varying Injection in GRB Afterglows}

\author{Re'em Sari$^1$ and Peter \Mesz$^{1,2,3}$}

\affil{$^1$ Theoretical Astrophysics 130-33, California Institute of Technology, Pasadena, CA, 91125\\
       $^2$ Institute for Advanced Study, Olden Lane, Princeton NJ 08540 \\
       $^3$ Pennsylvania State University, 525 Davey, University Park, PA 16802}

\begin{abstract}
The standard model of Gamma-Ray Bursts afterglows is based on synchrotron
radiation from a blast wave produced when the relativistic ejecta
encounters the surrounding medium. We reanalyze the refreshed shock
scenario, in which slower material catches up with the decelerating
ejecta and reenergizes it. This energization can be done either
continuously or in discrete episodes. We show that such scenario has
two important implications. First there is an additional component
coming from the reverse shock that goes into the energizing ejecta.
This persists for as long as the re-energization itself, which could extend
for up to days or longer. We find that during this time the overall spectral 
peak is found at the characteristic frequency of the reverse shock. 
Second, if the injection is continuous, the dynamics will be different from 
that in constant energy evolution, and will cause a slower decline of the observed
fluxes. A simple test of the continuously refreshed scenario is that it 
predicts a spectral maximum in the far IR or mm range after a few days.
\end{abstract}


\section{Introduction}

The standard model for GRB afterglows assumes that relativistic
material is decelerating due to interaction with the surrounding
medium.  A shock wave is formed heating the surrounding matter to
relativistic temperatures. It is assumed that both magnetic fields and
accelerated electrons aqcuire an energy density which is a significant
fraction of the equipartition value.  In the simplest case, which will
be referred to as the standard scenario, a single value of the energy
and the bulk Lorentz factor is injected either as delta or as a
top-hat function, of duration short respect to the afterglow. The
total energy is fixed in time and equals the initial energy of the
explosion.

Slower moving material is essential to all models which use density
gradients as a means of acceleration. Actually, if this is indeed the
mechanism, most of the system's energy is carried by the slower
material. This scenario overcomes the need for a clean
environment. The fact that one now needs a substantially higher energy
input can be addressed by a very energetic source such as a
massive star. A similar situation exists in some cases of
supernova. When a shock wave propagates through the envelope of the
star arrives at the edge it accelerates, and higher and higher velocities
are being imparted to a smaller fraction of the mass.

In GRB afterglows, if such slower material with significant energy is
ejected, it will affect the evolution in two major ways: first, the
system becomes more energetic as time passes (refreshed shock
scenario), therefore the temporal decay of the afterglow will be
slower (Rees \& \Mesz 1998).  Second, since the reverse shock will
last for as long as the energy supply continues, it adds an additional
long-living (reverse) emission component, typically at low
frequencies. The emission from such a reverse shock was considered
by \cite{kumpir99} for the discrete injection case. 
With accurate enough observations of the afterglow
temporal decay or good spectral sampling, especially at radio to mm
frequencies, both of these features may be detected, and could
therefore constrain the possibility of additional energy injection.

\section{Dynamics}

We assume here that the source ejects a range of Lorentz factors.  The
simplest description is that there is a certain amount of mass
$M(>\gamma )\propto \gamma ^{-s}$ moving with a Lorentz factor greater
than $\gamma $, all ejected at essentially the same time (
i.e., over a period which is
much shorter than the afterglow timescales). The energy associated
with that mass is $\gamma Mc^{2}\propto \gamma ^{-s+1}$.  Note that
this is valid only for $s>1$ , were for $s<1$ the energy above any
Lorentz factor is constant since it is all concentrated near the
highest Lorentz factor. We normalize this proportionality using the
initial Lorentz factor, $\gamma _{0}$, and the initial energy content,
which is about the ``burst'' energy $E_{0}$:
\begin{equation}
E(>\gamma )\sim E_{0}(\gamma /\gamma _{0})^{-s+1},
\label{eq:injection}
\end{equation}
down to some value $\gamma_{min} \leq \gamma_o$.  This tilted top-hat injection leads 
to a ``refreshed" shock scenario (\cite{rm98}), in contrast to the standard 
straight top-hat or delta function model with a mono-energetic $E_o$ and a 
single value $\gamma_o$.  Although $\gamma _{0}$ and $E_{0}$ are free 
parameters, we have lower limits on $\gamma _{0}\geq 100$ to avoid pair 
creation during the GRB itself, and we have an estimate of the initial 
energy $E_{0}$ from the energy 
seen in the burst. The actual value of $\gamma _{0}$ can be obtained from the 
onset time of the afterglow or from the reverse shock initial frequency. 
This scenario is related to, but not identical, to the \cite{fenram99} 
model in which a wind with near-steady $\gamma$ impacts on a decelerating 
'wall' produced by the outermost shells of material which first make contact
with the exterior gas. The refreshed shock or tilted top-hat scenario envisages 
an 'engine' duration which is instantaneous compared to the deceleration time,
whereas the latter scenario assumes that the wind duration is longer than
the initial deceleration time.  However if the 'engine' produced a wind whose 
Lorentz factor decreased with time, 
the net effect could be rather similar to the refreshed shock.

The low Lorentz factor mass will catch up with the high Lorentz factor
mass only when the latter has decelerated to a comparable Lorentz
factor. At that time the shocked material (both reverse and forward
shock) has a Lorentz factor $\gamma $ satisfying
\begin{equation}
E_{0}(\gamma /\gamma _{0})^{-s+1}\sim E\sim \gamma ^{2}R^{3}\rho c^{2}.
\label{eq:gamma}
\end{equation}

If we assume that the outer density is $\rho \propto R^{-g}$, then we have 
\begin{equation}
\gamma =\gamma _{0}(R/R_{0})^{-(3-g)/(1+s)}
\end{equation}
Using the relation $t\sim R/\gamma ^{2}c$ we get 
\begin{eqnarray}
R &=&R_{0}(t/t_{0})^{(1+s)/(7+s-2g)}  \label{eq:Rscalings} \\
\gamma &=&\gamma _{0}(t/t_{0})^{-(3-g)/(7+s-2g)}~, \nonumber
\end{eqnarray}
where $R_0$ and $t_0$ are the deceleration radius and deceleration time of
the initial material with $E_0,~\gamma_O$.
Note that these scalings differ from those of Rees and \Mesz (1998),
who obtained $s+1$ where we have $s$ in the scalings of expressions
after our equation (\ref{eq:gamma}). For $s=1$ the energy in the slow
material adds up only logarithmically and therefore the above
expression degenerates to the usual ones of instantaneous
injection. However, for $s>1$ where more energy is stored in a slowly
moving material, we have a slower decay as more and more energy is
added to the system as time evolves. Note also that these expression
are only valid for $g<3$. Otherwise, the shocks is accelerating (see
Blandford \& Mckee for $3<g<4$ and Best \& Sari for $g>4$)$.$ However,
the most useful values of $g$ are probably the constant density case
$g=0$ and the wind case $g=2$.

Since the additional slow shells catch up with the shocked material
once these are with comparable Lorentz factors, the reverse shock is
always mildly relativistic (Rees \& \Mesz 1998, Kumar and Piran
1999). The thermal Lorentz factor of the electrons is therefore
roughly given by the ratio of proton to electron mass $\gamma
_{e}^{r}\sim \epsilon _{e}m_{p}/m_{e}$. In the forward shock, the
thermal Lorentz factor of the electrons is $\gamma _{e}^{f}\sim
\epsilon _{e}\gamma m_{p}/m_{e}$.

The pressure behind the reverse shock is proportional to the density
behind the reverse shock which is comparable to the density in front
of it (a mildly relativistic shock) therefore
\[
p_{r}\propto n_{r}\propto \frac{M}{R^{3}/\gamma }\propto t^{-\frac{6+sg-g}{%
7+s-2g}}. 
\]
As a check we can see that this is also the pressure at the forward
shock which is proportional to $\gamma ^{2}\rho $.  The forward and
reverse shock have the same bulk Lorentz factor and the same pressure,
while the forward shock has a temperature which is higher by a factor
of $\gamma$.

\section{Radiation}

The dynamics above determine the bulk Lorentz factor and the thermal
Lorentz factor of the electrons as function of time. To estimate the
resulting synchrotron radiation, one also needs an estimate of the
magnetic field.  For a given magnetic field, the spectrum consists of
four power law segments separated by three break frequencies $\nu
_{a}$, $\nu _{m}$ and $\nu _{c}$ (Sari, Piran and Narayan 1998, \Mesz,
Rees \& Wijers 1998). Adopting the standard assumption that the
magnetic energy density is some fraction $\epsilon _{B}$ of
equipartition, i.e., proportional to the pressure, we have
\[
B\propto t^{-\frac{1}{2}\frac{6+sg-g}{7+s-2g}}. 
\]
Since the pressure in the reverse and forward shock is identical, the magnetic field 
will also be the same, if the equipartition parameter $\epsilon_B$ is the same. 
The difference between the forward and 
reverse shock lies then in the number of electrons 
(which is larger by a factor of $\gamma$ at the reverse shock) and their
thermal Lorentz factor (which is smaller by a factor of $\gamma$ at
the reverse shock). This result in the following general
properties, valid at any given moment:

1) The peak flux of the reverse shock, at any time, is larger by a factor of 
$\gamma $ than that of the forward shock: $F_{\nu ,\max }^{r}=\gamma F_{\nu
,\max }^{f}$

2) The typical frequency of the minimal electron in the reverse shock is
smaller by a factor of $\gamma ^{2}$: $\nu _{m}^{r}=\nu _{m}^{f}/\gamma ^{2}$.

3) The cooling frequency of the reverse and forward shock are equal: 
$\nu_{c}^{r}=\nu _{c}^{f}=\nu_c$.

4) At sufficiently early time (typically the first few weeks or
months) $\nu _{a}^{r,f}<\nu _{m}^{r,f}$ and $\nu _{a}^{r,f}<\nu
_{c}$. The sefl absorption frequency of the reverse shock is larger than
that of the forward shock. It is larger by a factor of $\gamma ^{3/5}$
initially, when both are in fast cooling and by a factor of $\gamma
^{8/5}$ if both are in slow cooling.

Points one and two above agree with those of Kumar and Piran, calculated 
for the discrete case. We have here generalized the result to include the
effect of the cooling frequency and self absorption frequency.
The combined reverse+forward shock emission can therefore be one of three
types, evolving in time in the following order:

A) Both reverse and forward shock are cooling fast:
$\nu_{a}^f<\nu_a^r<\nu _{c}<\nu_{m}^{r}<\nu _{m}^{f}$.

B) Reverse is slow cooling, forward is fast cooling: $\nu_{a}^f<\nu_a^r<\nu _{m}^{r}<\nu _{c}<\nu _{m}^{f}$.

C) Both reverse and forward shock are in slow cooling,
$\nu_{a}^f<\nu_a^r<\nu _{m}^{r}<\nu_{m}^{f}<\nu _{c}$.

\begin{figure}
\begin{center}
\epsscale{0.75}
\plotone{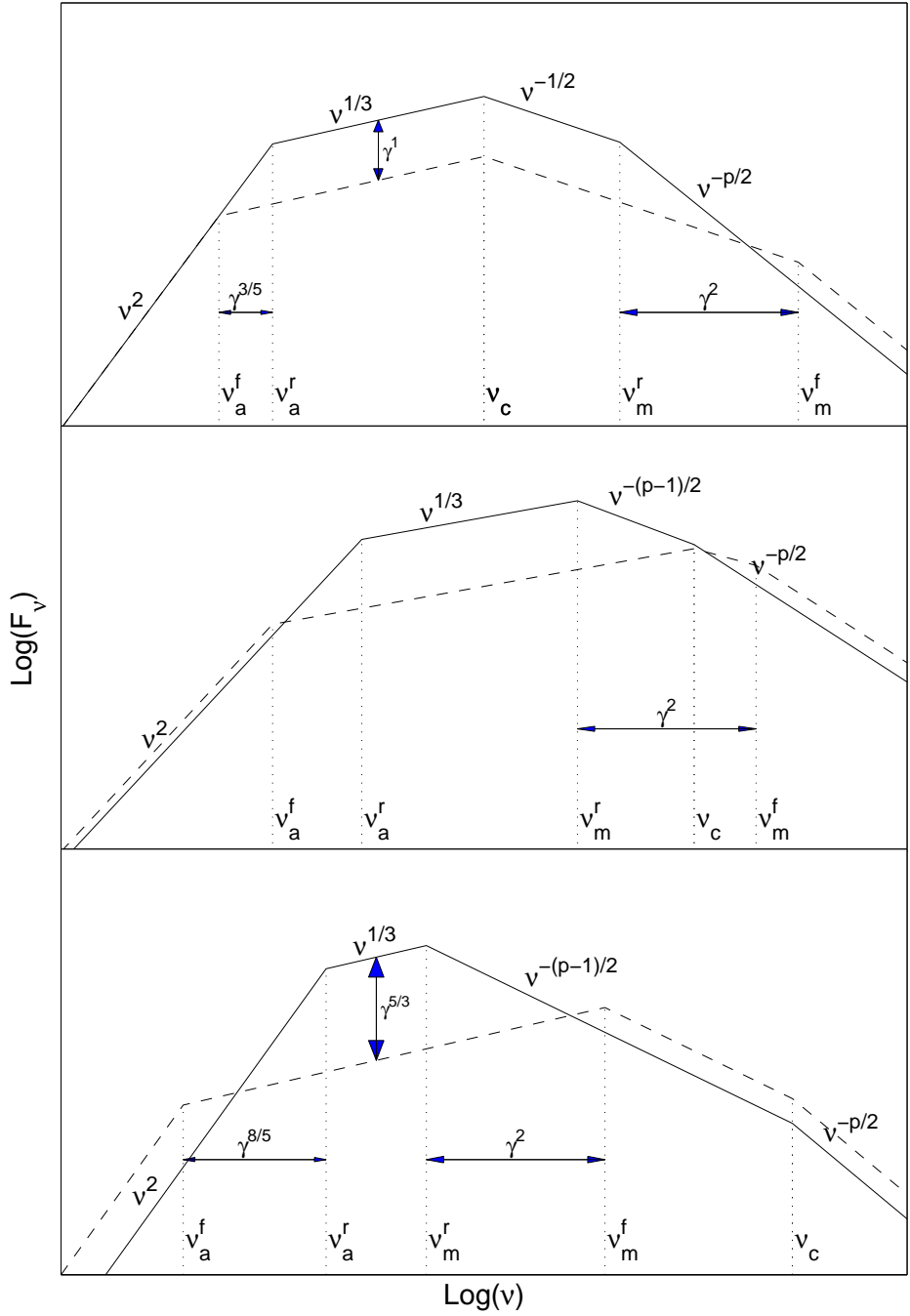}
\caption{Spectrum of the reverse and forward shocks with synchrotron peaks at
$\nu_m^r$ and $\nu_m^f$, synchrotron self-absorption frequencies $\nu_a^r$
and $\nu_a^f$ and cooling frequency $\nu_c$ (same for both, assuming
$\epsilon_B^f=\epsilon_B^r$), for an electron injection spectrum $\propto
\gamma^{-p}$. Top: both shocks fast cooling. Middle: reverse slow, forward
fast cooling. Bottom: both slow cooling.}
\label{spectra}
\end{center}
\end{figure}

These three spectra are presented in figure \ref{spectra}.  As evident
from the figure, the forward shock dominates the emission at very low
and very high frequencies while the reverse shock contributes to a
spectral ``bump'' at intermediate frequencies.  The peak flux is that
of the reverse rather than the forward shock. The combined spectrum is
somehwat flatter than the usual one (which uses the forward shock
only). A good sampling of the spectrum, especially at low frequencies,
can therefore show the existence or non-existence of such a feature.
The forward shock alway dominates above $\nu>\max(\nu_m^f,\nu_c)$ by a
small factor of $\gamma^{p-2}$. Since the value of $p$ is close to
$2$, the forward shock does not radiate much more than the reverse at
high frequencies.

\bigskip
{\it -- FIGURE 1 --}
\bigskip

In the case of fast cooling we have ignored the effect of the ordered
structure of the electron's energy behind the shock (\cite{gran00}), both
for the reverse and forward shock. This effect will increase the
emission at frequencies below the self absoption frequency,
$\nu<\nu_a$, but will not change the qualitative conclusions of this
paper.

The spectrum displayed in figure \ref{spectra} is valid at any moment
if the energy and momentum injection is continuous, but also at the
moment of impact in the case of the standard top-hat injection. However, in 
the latter case the reverse shock component will rapidly disappear as discussed 
by Sari \& Piran 1999 and \Mesz \& Rees 1999. If the injection is discrete, the 
dynamics of the forward shock right after the collision will not be affected,
and it will evolve as in the standard non-refreshed scenario.
.

For the continuous case, the time dependence $t^{-q}$ of the various
quantities is given in Table 1, for arbitrary parameters $s$ and $g$,
assuming a spectral shape $\propto \nu^{-\beta}$. Above the peak
$\nu_{max}=\min [\nu_m,\nu_c]$ where the flux has the value $F_{\nu,max}$
the dependence $F_\nu\propto t^{-\alpha}\nu^{-\beta}$ is calculated
separately for the slow and fast cooling regimes.

\bigskip
{\it -- TABLE 1 --}
\bigskip

To give more specific numerical examples we specialize to the constant
density case, where $g=0$. We then have
\begin{eqnarray}
\nu_m^f=2.0 \times 10^{13} {\rm Hz\,} (1+z)^{1/2} \epsilon_{B,-2}^{1/2} 
\epsilon_{e,0.5}^2 E_{52}^{1/2} t_{day}^{-3/2}
(\frac t {t_0})^{\frac{3(s-1)}{2(7+s)}} \\
\nu_m^r=9.1 \times 10^{11} {\rm Hz\,} (1+z)^{-1/4} \epsilon_{B,-2}^{1/2} 
\epsilon_{e,0.5}^2 E_{52}^{1/4} n_0^{1/4} t_{day}^{-3/4} (\frac t {t_0})^{\frac{3(s-1)}{4(7+s)}} \\
\nu_c=2.7 \times 10^{15} {\rm Hz\,} (1+z)^{-1/2} \epsilon_{B,-2}^{-3/2} E_{52}^{-1/2} 
n_0^{-1} t_{day}^{-1/2} (\frac t {t_0})^{-\frac{3(s-1)}{2(7+s)}} \\
F_{\nu,max}^f=2.6 {\rm mJy\,} (1+z) \epsilon_{B,-2}^{1/2} E_{52} n_0^{1/2} 
D_{L,28}^{-2} (\frac t {t_0})^{\frac{3(s-1)}{7+s}} \\
F_{\nu,max}^r=12 {\rm mJy\,} (1+z)^{11/8} \epsilon_{B,-2}^{1/2} E_{52}^{9/8} n_0^{3/8} 
D_{L,28}^{-2} t_{day}^{-3/8} (\frac t {t_0})^{\frac{27(s-1)}{8(7+s)}}
\label{eq:fnumax}
\end{eqnarray}
%where $\nu_{max}=\min[\nu_c,\nu_m]$ is the frequency at which the spectrum
%peaks. 
For slow cooling, $\nu _{c}>\nu _{m}>\nu _{a}$, the spectral peak is
at $\nu_{max}=\nu_m$, and synchrotron self-absorption occurs at
\begin{eqnarray}
\nu _{a}^{f} = 3.6{\rm GHz\,} (1+z)^{-1} \epsilon_{e,0.5}^{-1} \epsilon_{B,-2}^{1/5}
E_{52}^{1/5} n_0^{3/5} (\frac t {t_0})^{\frac{3}{5} \frac{s-1}{7+s}} \\
\nu_a^r=43 {\rm GHz\,} (1+z)^{-2/5} \epsilon_{e,0.5}^{-1} \epsilon_{B,-2}^{1/5} E_{52}^{2/5} n_0^{2/5} t_{day}^{-3/5} (\frac t {t_0})^{\frac{6(s-1)}{5(7+s)}}
\end{eqnarray}
while for fast cooling, $\nu _{m}>\nu _{c}>\nu _{sa}$, the spectral peak 
is at $\nu_{max}=\nu_c$ and we have
\begin{eqnarray}
\nu _{a}^{f} = 0.3{\rm GHz\,} (1+z)^{-1/2} \epsilon_{B,-2}^{6/5}
E_{52}^{7/10} n_0^{11/10} t_{day}^{-1/2} 
(\frac t {t_0})^{\frac{21}{10} \frac{s-1}{7+s}} \\
\nu_a^r=0.8 {\rm GHz\,} (1+z)^{-11/40} \epsilon_{B,-2}^{6/5} E_{52}^{31/40} n_0^{41/40} t_{day}^{-29/40} (\frac t {t_0})^{\frac{93(s-1)}{40(7+s)}}
\end{eqnarray}

Using the normalization of the peak flux and the break points, the
flux can be calculated at any frequency. Similar to the standard case,
it is possible to test the model by comparing the temporal decay and
spectral slopes.

In the standard case (mono-energetic instantaneous injection), the
flux above this frequency is falling with time, while the flux
below this frequency is rising with time. The additional energy in the
varying injection case tends to flatten the decay rate, and for high
enough values of $s$ can even make it grow. 
(From equations (1) and (4) one sees that the $t/t_0$ factors in equations (5)-(13)
are equivalent to a power of the ratio of  the injected to initial energy $E/E_0$).
Stated differently, one
would need a steeper spectral index to give rise to the same observed
temporal decay in the refreshed scenario. Table 2 summarizes for $g=0$
the values of the spectral index $\beta$ that can be inffered from a
measured temporal decay index $\alpha$ in the instantenous and
refreshed scenarios, for reverse and forward shocks.

\bigskip
{\it -- TABLE 2 --}
\bigskip

It can be seen that the spectal indices that need to exmplain a
$t^{-1}$ decay that is observed in many bursts are considerably
steeper. Some confusion can occur between a forward 
moderately refreshed ($s=2$) shock in
the slow cooling regime and a fast cooling forward shock in the
non-refreshed scenario, as these two scenarios predict similar
relation between $\alpha$ and $\beta$ for nominal values.
However, most other regimes are considerably different from 
the standard instanteneous forward shock prediction even if 
one only has moderately accurate spectral data information.

\section{Specific Bursts}

\noindent{\it GRB 970508}. - In the case of GRB 970508, the
observations show a steep increase in the optical and X-ray fluxes
between one and two days. This can be interpreted in terms of a
varying injection event, e.g. Panaitescu, \Mesz \& Rees 1998, who
consider the radiation of the forward shock assuming a large value of
$s$, a value of $\gamma_{min}=11<\gamma< \gamma_o$ and a final energy
$E_f=3E_o$. After this time the energy is constant, and the afterglow
can be fitted with a standard mono-energetic afterglow, e.g. Wijers \&
Galama 1999.  How could we test the hypothesis that the "jump" between
half a day to two days is indeed due to varying injection? It turns
out that a delayed energy injection (if $\epsilon_{B}$ is the same for
the reverse and forward shocks) has a very strong prediction. We can
estimate the forward shock break frequencies at $t=2$ days by
extrapolating them back from those at day 12, where the spectrum is
well studied (e.g. \cite{wiga99}, \cite{gran99b}). 
This results in $\nu_m^f \cong 1.3
\times 10^{12}$ Hz, $F_{\nu,max}\sim 1.7$mJy, $\nu_a^f \cong 3$ GHz and
$\gamma=4$. Therefore, according to the lower frame of figure
\ref{spectra} the reverse shock should have $\nu_m^r\cong 80$GHz and
$\nu_a^r\cong 30$GHz.  The reverse shock signature should be the
largest between these two frequencies, where the
flux should exceed by an order of magnitude the simple extrapolation
of the forward shock model back to day two. At a more observationally
accessible frequency of $20$GHz and $100$GHz the flux increase due to
the reverse shock should be a factor $\sim 5$ and $\sim 8$
respectively, i.e. a flux of $2.2$mJy and $6$mJy, respectively.
Unfortunately, the observations around 2 days at these frequencies
where short and therefore of low sensitivity. The $3\sigma$ upper
limit of $6$mJy obtained by BIMA is consistent with energy injection.
This prediction of an additional low frequency component in GRB~970508 is
similar to that of \cite{kumpir99}. However, taking the self absorption 
frequency into account we have shown here that this would not apply to the
usually observed radio frequencies $1.4$GHz-$8.4$GHz but to the range
of $30$GHz-$80$GHz.
It is therefore of great value to obtain low frequency observations
or strong upper limits if such jumps in the optical and X-ray flux
will be seen again in the future.  If the reverse shock signature is
not seen in other comparable bursts where a light curve peak is
detected at a certain time (1.5 days in this case), a varying
injection episode can be ruled out as an explanation for this peak (or
else, it would imply $\epsilon_{B,rev} \ll \epsilon_{B,for}$).
%The temporal decay index after 2 days is $\alpha=1.2$ (CHECK?) and the
%spectral index $\beta\simeq 0.75$ (Wijers \& Galama 1999), which indicates
%that after 2 days the decay is well fit by a monoenergetic standard model
%with (slow cooling case) $\alpha=(3/2)\beta$, which means that if injection
%was responsible for the step at 1-2 days, the tilted top-hat injection spectrum 
%of equation (\ref{eq:injection}) is limited to values $\gamma_{min} \leq
%\gamma \leq \gamma_o$ such that the material with $\gamma_{min}$ (estimated
%as $\gamma_{min}\simeq 11$ in Panaitescu \etal 1998) arrives at the shock 
%surface at $t\sim 1.5$ days.

\noindent{\it GRB 990123}. - In GRB 990123, a bright 9th magnitude
prompt flash was seen about 60s after the trigger time (\cite{ake99}),
attributable to a reverse shock (\cite{sp99a,mr99} with an initial
temporal decay $F\propto t^{-2}$, steeper than the subsequent decay
$F\propto t^{-1.1}$ which is attributable to the forward shock. Here,
an impulsive mono-energetic (standard) injection fits better than a
tilted top-hat, since e.g a varying $s=2,g=0$ injection would predict
a decay $\alpha=1$, while an impulsive event gives $\alpha=2$, in good
agreement with observations for a spectral slope $\beta=1$ in the fast
cooling regime (the values are 2/3 and 9/8 in the adiabatic regime).
Though no spectral information was available for this burst in the
first hours, a slope $\beta\le 1$ can be assumed based on our
accumulated knowledge from previous afterglows.

\section{Discussion}

The dynamics and emission of the forward and reverse shocks is controlled
by several factors, including the continuity and nature of the energy
and mass input, the possible existence of external density gradients,
and the strength of the magnetic fields in these regions.  
A contineous injection of
energy with a lower Lorentz factor has as its main consequence that it
tends to flatten the decay slopes of the afterglow, after it has gone
through the maximum. An external density gradient (e.g. as in a
wind with $\rho_{ext} \propto r^{-g}$, where $g\sim 2$ might be
typical) has the property of steepening the decay. This applies in
general both to the forward and the reverse shocks.  

We have suggested ways in which one can attempt to discriminate
between the standard straight top-hat injection of energy and momentum
with a single $\gamma_o$ and $E_o$, which then remains constant
throughout the afterglow phase, and a refreshed scenario, where the
injection is also brief (e.g. comparable to the gamma-ray burst
duration and therefore instantaneous compared to the afterglow
timescale) but in which there is a varying distribution of $\gamma$
and of energy during that injection, so that matter ejected with low 
Lorentz factor cateches up with the bulk of the flow on long timescales.
The afterglow energy then increases with time. 
Under the simple assumption that both the
forward and the reverse magnetic fields are equal
($\epsilon_{B,r}=\epsilon_{B,f}$) a remarkable prediction is that in
all regimes (both shocks are fast cooling, 
reverse is slow cooling and forward shock
is fast cooling or both are slow cooling) the
reverse shock spectrum joins seamlessly, or with only a very modest
step $\propto \gamma^{p-2}$, onto the forward shock spectrum,
extending it to lower frequencies. (This could be modified if, for
instance, $\epsilon_{Br} \ll \epsilon_{Bf}$, which would give a
spectrum with a more pronounce through separating the reverse and
forward components).  Specifically, in the case of GRB 970508, if a
descrete episode of injection that produced refreshed shocks at 
$t\sim 1.5$ days is
the explanation of the step in the X-ray and optical flux at this
time, then one would expect (for equal forward and reverse
$\epsilon_B$) an even more dramatic rise of the  20 GHz  and 100GHz flux 
at 1.5-2 days. This should be about a factor of 5 and 8, respectively, 
larger than expected from the forward shock values extrapolated back
to 2 days. In the case of GRB 990123 a reverse shock appears to
have been responsible for the prompt optical flash, and the decay
indices (for reasonable spectral slopes) are compatible with
monoenergetic impulsive (or straight top-hat) injection.

For future GRB afterglow observations, the main prediction from having
comparable values of $\epsilon_B$ in the forward and reverse shocks of
baryon loaded fireballs is that 
the peak flux is found at the peak
frequency of the reverse, rather than of the forward shock, i.e. at
lower frequencies than typically considered. The IR, mm and radio
fluxes would therefore be expected to be significantly larger than for
simple (forward shock) standard afterglow models (e.g. Fig 1).  
This holds whether the injection in contineous or descrete.
The two contributions continue to evolve as a pair of smoothly joined
components, the ratio of the two peak frequencies
$\nu_{m}^r/\nu_{m}^f\propto \gamma^{-2}$ and peak fluxes
$F_{\nu,mr}/F_{\nu,mf} \propto \gamma$ gradually approaching each
other until they coincide at the transition to the non-relativistic
case $\gamma\sim 1$.

\acknowledgements We are grateful to P. Kumar, A. Panaitescu, 
T. Piran and M.J. Rees for comments.  
RS is supported by the Sherman Fairchild foundation. 
The research of PM is supported through NASA NAG 5-2857, 
the Guggenheim Foundation, the Division of Physics, Math \& Astronomy,
Astronomy Visitor Program and the Merle Kingsley fund at Caltech, and
the Institute for Advanced Study.

\eject

% NEW TABLE 1

\begin{table*}
\begin{center}
\def\frh{}
\def\dn{2(7+s-2g)}
\begin{tabular}{cccccccccc}
\tl \tl
 & $\nu_m$ & $ F_{\nu_m}$ & $\nu_c$ & $F_\nu$:~$\nu_m<\nu<\nu_c$  & $F_\nu$:~$\nu>\max(\nu_c,\nu_m)$ \cr
 
\tl 

f & -$\frh\frac{24-7g+sg}{\dn}$   & $\frh\frac{6s-6+g-3sg}{\dn}$ & 
          -$\frh\frac{4+4s-3g-3sg}{\dn}$& -$\frh\frac{6-6s-g+3sg+\beta(24-7g+sg)}{\dn}$ & -$\frh\frac{-4-4s+g+sg+\beta(24-7g+sg)}{\dn}$ \cr
\tl
%for & "~~~ & "~~~ & "~~~ & $-\frh\frac{-4-4s+g+sg+\beta(24-7g+sg)}{\dn}$ \cr
%\tl
r &  -$\frh\frac{12-3g+sg}{\dn}$   & $\frh\frac{6s-12+3g-3sg}{\dn}$ & 
        -$\frh\frac{4+4s-3g-3sg}{\dn}$& -$\frh\frac{12-6s-3g+3sg+\beta(12-3g+sg)}{\dn}$ & -$\frh\frac{8-4s-3g+sg+\beta(12-3g+sg)}{\dn}$ \cr
\tl
%rev & "~~~ & "~~~ & "~~~ & $-\frh\frac{8-4s-3g+sg+\beta(12-3g+sg)}{\dn}$ \cr
\tl
\end{tabular}
\tablenum{1}
\caption{ Temporal exponents of the peak frequency $\nu_m$, the maximum flux 
$F_{\nu_m}$, the cooling frequency $\nu_c$ and the flux in a given bandwidth 
$F_\nu$, calculated both in the adiabatic regime $\nu_m < \nu <\nu_c$ 
($F_\nu\propto F_{\nu_m}(\nu_m/\nu)^\beta \propto t^{-\alpha}\nu^{-\beta}$, 
where $\beta=(p-1)/2$), and in the cooling regime $\nu_c<\nu_m<\nu$ 
($F_\nu\propto(\nu_c/\nu_m)^{1/2}(\nu_m/\nu)^\beta \propto
t^{-\alpha}\nu^{-\beta}$ where $\beta=p/2$). }
\end{center}
\end{table*}
%

% NEW TABLE 2

\begin{table*}
\begin{center}
\begin{tabular}{|c||c||cc|cc|cc|}
\tl 
        &                           & \multicolumn{6}{|c|}{ spectral index $\beta$, as function of $\alpha$ ($F_\nu \propto t^{-\alpha}\nu^{-\beta}$)}                \cr
\raisebox{1.5ex} {Shock}   &  \raisebox{1.5ex} {Regime}                   & \multicolumn{2}{|c|}{ no-injection}                & \multicolumn{2}{|c|}{$s=2$}                 & \multicolumn{2}{|c|}{$s=3$}                 \cr \tl
forward & $\nu_m<\nu<\nu_c$         & $\frac{2\alpha}{3}$   &         $\left[2/3\right]$ & $\frac{3\alpha+1}{4}$ & $\left[1\right]$    & $\frac{5\alpha+3}{6}$  & $\left[4/3\right]$   \cr \tl 
forward & $\nu>\max({\nu_m,\nu_c})$ & $\frac{2\alpha+1}{3}$ & $\left[ 1 \right]$         & $\frac{3\alpha+2}{4}$ & $\left[5/4\right]$  & $\frac{5\alpha+4}{6}$  & $\left[3/2\right]$  \cr \tl
reverse & $\nu_m<\nu<\nu_c$         & \multicolumn{2}{|c|}{short-lived}                  & $\frac{3\alpha  }{2}$ & $\left[3/2\right]$  & $\frac{10\alpha+3}{6}$ & $\left[13/6\right]$  \cr \tl
reverse & $\nu>\max({\nu_m,\nu_c})$ & \multicolumn{2}{|c|}{short-lived}                  & $\frac{3\alpha  }{2}$ & $\left[3/2\right]$  & $\frac{5\alpha+1}{3}$  & $\left[2 \right]$  \cr \tl
\end{tabular}
\tablenum{2}
\caption{ The spectral indices $\beta$ that would be deduced from an
observed temporal decay index $\alpha$. The third column gives the
well known ``standard model'' results for instantenous injection for
which the reverse shock does not last. The fourth and fifth columns
are for contineous injection case with $s=2$ and $s=3$
respectively. For extended emission the reverse shock lives as long as
the slower material keeps arriving, e.g. days. The value in brackets
demonstrate the numerical value for a nominal temporal index of
$\alpha=1$.  }
\end{center}
\end{table*}

\eject

\end{document}